\documentclass[conference, a4paper,UTF8]{IEEEtran}
\IEEEoverridecommandlockouts
\usepackage{fancyhdr} 
\usepackage{color}
\usepackage{amssymb}
\usepackage{amsmath}
\usepackage{amsmath,amssymb,amsfonts}
\usepackage{algorithmic}
\usepackage{graphicx}
\usepackage{textcomp}
\usepackage{amsmath,amssymb,amsfonts}
\usepackage{algorithmic}
\usepackage{graphicx}
\usepackage{float}
\usepackage{subfigure}
\usepackage{setspace}
\usepackage{hyperref}
\usepackage{textcomp}
\usepackage{amsfonts}
\usepackage{paralist}
\usepackage{tikz}
\usetikzlibrary{shapes.arrows}
\usepackage{amsmath}
\usepackage{algorithm}
\usepackage{algorithmic}
\usepackage{amsthm}
\usepackage{warpcol}
\usepackage{booktabs}
\usepackage{multirow}
\usepackage{graphicx}
\usepackage{booktabs}
\usepackage{rotating}
\usepackage{placeins}
\usepackage{mathrsfs}
\usepackage{ifsym}
\usepackage{tikz} 
\usepackage{subfigure}
\usepackage{geometry}
\usetikzlibrary{decorations.pathreplacing}
\usetikzlibrary{decorations.pathreplacing,calligraphy}
\usepackage{makecell}
            
\usepackage{lastpage}      
\usepackage{setspace}          
\usepackage{algorithm} 
\usepackage{algorithmic} 
\usepackage{multirow} 

\usepackage{amsmath,bm} 
\usepackage{xcolor}                         
\usepackage{layout}  
\usepackage{amsmath}   
\usepackage{hyperref}
\usepackage{color}
\usepackage{amsfonts}
\usepackage{graphicx}       
\usepackage{hyperref}                                  
\begin{document}
\begin{spacing}{1}

\title{The Electromagnetic Balance Game: A Probabilistic Perspective}
\author{
\IEEEauthorblockN{Fangqi Li}
\IEEEauthorblockA{\textit{School of Cyber Science and Engineering, SEIEE, SJTU}\\
\IEEEauthorblockN{$\{$solour\_lfq$\}$@sjtu.edu.cn}}
}
\date{Today}
\maketitle
\begin{abstract}
Finding a counterfeit coin with the different weight from a set of visually identical coin using a balance, usually a two-armed balance, known as the balance question, is an intersting and inspiring question. Its variants involve diversified toolkits including information theory, coding theory, optimization, probabilistic theory, combinatorics and a lot of quick wits. In this paper some variants of the balance game are dicussed, especially from a probabilistic perspective. Unlike the gravity field setting, we adopt an electromagnetic field, where tighter bounds for some variants of the balance game can be found. We focus on the predetermined setting, where the player has to arrange the strategy without observing the outcome of the balancing. The sufficient condition for the balance to win is obtained by adopting a coding scheme.  Apart from designing a delicate encoding framework, we also propose and analyze the performance of a completely randomized strategy. The optimal behavior of a randomized player is derived. Then we rise the dishonest balance game, in which the balance can adversely cheat the player. We present some elementary results on the analysis of dishonest balance game using probabilistic method at length. Its relationship with Shannon' s coding theorem in a noisy channel is also revealed. 


\end{abstract}
\section{Introduction}
\label{section:1}
The \emph{balance question} appears as an IQ test as well as a good point from which one begins an introductory lecture on information theory. A simple balance question often involves a fair balance and a set of visually identical coins in which only a counterfeit one has a different weight from the others. For example: 

\emph{There is one heavier counterfeit coin among twelve coins, the rest eleven ones weigh equally. Given a two-armed balance, at least how many times does it take to find the overweight coin?}

Elements from information theory such as entropy provide an elegant bound to this question. Balance question has many variants, e.g., we might only know that one coin has a different weight (but we do not  know whether it is heavier or lighter), the number of coins could be an arbitrary integer, the balance might be biased, etc. 

Some late variants adopt even more far-fetching assumptions such as there are more than one counterfeit coins, the balance is multi-armed, etc \cite{de1998predetermined,wen2004optimal,liu2005searching}.

The traditional balance setting in a gravity field often asserts that the numbers of coins on two sides of the balance are equal. We adopt an electromagnetic setting where the genuine coins are neutral while the counterfeit coin is electrified. So the numbers of coins on two sides of the balance need not to be identical. This setting yields some tighter bounds than the gravity setting. 

In the balance question, we consistently play as the human player who tries to find the counterfeit coin. What if the player has to determine his/her strategy beforehand instead of deciding it adaptively? In \emph{balance game} one takes the position of the balance and tries to hide the counterfeit coin from the human player. The same set of methods can be applied to the game setting parallely. After reviewing the traditional understanding of the balance question, we study the \emph{balance game}. Finally, we introduce the \emph{dishonest balance game}, in which the balance can cheat the human player. We present some analysis over this game. 

The contributions of this paper are:
\begin{enumerate}
\item We propose the balance game in an electromagnetic setting. A coding framework, together with probabilistic method is adopted to analyze the balance question/game under this setting.
\item Some results on the balance game are derived under this framework. Such as the winning condition for the human player or the balance.  Moreover, we prove an interesting result: when the player adopts a complete stochastic strategy, it is optimal for him/her to put any coin on the left/right side of the balance or off the balance independently and uniformly (each with probability $\frac{1}{3}$) at each round. 
\item The dishonest balance game is proposed and an elementary bound on the winning strategy of the balance is provided.
\end{enumerate}

Section \ref{section:2} covers a review on the traditional understanding of the balance question from information entropy. Section \ref{section:3} introduces the coding formulation with probabilistic method on the balance question as well as the balance game. Section \ref{section:4} introduces the dishonest balance game, together with some games that motivate this proposal, an analytic bound on the dishonest balance game is given. Section \ref{section:5} gives the conclusion and some discussions. 

\section{Balance Question and Entropy}
\label{section:2}
To quantify a balance question/game, we use $(n,q,\text{prior})$ to indicate the setting, where $n$ is the number of coins, $q$ the number of measuing and the third parameter denotes the prior information. 

It is known that \emph{entropy} is a good metric in measuring the quantity of information \cite{cover1999elements}. 
One lower bound for the number of human player's moves in balance question is obtained using entropy. Assuming that one wants to find one overweight coin out of $n$ coins. There are altogether $n$ possibilities (each of the $n$ coins could be heavier), so the entropy of this set of coins is no higher than $\log_{2}n$ bits. Each time the balance yields a result, the information released is at most $\log_{2} 3$ bits (the balance can tilted to either side or stay unbiased). Thus in the worst case, the player is impossible to detect the correct coin with less than
$$\frac{\log_{2}n}{\log_{2}3}=\log_{3}n$$
balancings. That is to say, if the human player is not allowed to use the balance for more than $\log_{3}n-1$ times then it is always possible (for the balance) to hide a coin such the player cannot find it. Generalization to a special coin with either higher/lower weight is similiar, one simply replace the number of possibilities from $n$ to $2n$. However, even if the player is allowed to use the balance for $\log_{3}2n$ times, he/she is not guaranteed to correctly find the counterfeit coin. The reason behind is that the player might fail to design a configuration of coins so that the probabilities for the balance to yields three different results are not identical, and the information yields by each balancing is less than $\log_{2} 3$, hence the total amount of information is insufficient. Occasionally, there are cases where the discrete nature of balancing is contradictive to the theory. This is often observed when $n$ is small. For example, if the player is asked to find one coin with different weight from $n$ coins with $q=3$ times of balancings, then entropy guarantees that $n\leq 13$. But in practice we must have $n \leq 12$. That is to say, the player can never win a $(13,3,\text{unknown})$-balance question. To see this, we write the space of all possibilities as $H=\left\{1^{-},1^{+},\cdots,13^{-},13^{+} \right\}$, where $i^{+}(i^{-})$ means that the $i$-th coin is heavier(lighter). During the first balancing, the human player can do not more than putting $a$ coins on both side of the balancing (w.l.o.g. we assume that the player puts coins indexed 1 to $a$ on the left, those indexed $a+1$ to $2a$ on the right), which is going to divide $H$ into three partitions:
$$\left\{1^{+},2^{+},\cdots,a^{+},(a+1)^{-},\cdots,(2a)^{-}\right\},$$
$$\left\{1^{-},2^{-},\cdots,a^{-},(a+1)^{+},\cdots,(2a)^{+}\right\},$$
$$\left\{(2a+1)^{+},(2a+1)^{-},\cdots,13^{+},13^{-} \right\}.$$
Whose sizes are $2a,2a,26-4a$. To make sure the next two balancings can yield an answer, it is necessary that $2a\leq 9$ and $26-4a \leq 9$, resulting in:
$$4.25 \leq a\leq 4.5.$$
Which is inconsistent with the fact that $a$ must be an integer. But when the number of chips $n$ grows very large, it is almost always safe to expect that $\log_{3} n$ times of balancing can yield a good result. 

\section{Balance Game}
\label{section:3}
Now we turn to the balance's position and study the \emph{balance game}. In the balance question, the human player plays with an \emph{adaptive strategy},  i.e., the configuration of coins at the $j+1$-th round is decided given the tilting situation of the previous $j$ rounds. To design a strategy for the balance, we now have the human player design a \emph{predetermined strategy} without observing the balance. At the beginning of the game, the human player has to determine which coins be put onto the left/right side of the balance or off the balance at all $q$ rounds. The balance then yields a sequence of weighing results from which the human player tries to deduce the index of the exceptional coin. 

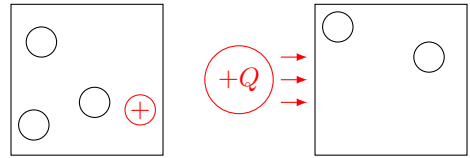
\begin{figure}[htb]
\centering
\begin{tikzpicture}
\draw (0,0) rectangle (2,2);
\draw (3,1) [red] circle (0.45);
\node at (3,1) [red] {$+Q$};
\draw (4,0) rectangle (6,2);
\draw (0.3,0.4) circle (0.2);
\draw (1.1,0.7) circle (0.2);
\draw (0.4,1.5) circle (0.2);
\draw (1.7,0.6) [red] circle (0.2);
\node at (1.7,0.6) [red] {$+$};
\draw (5.5,1.3) circle (0.2);
\draw (4.3,1.7) circle (0.2);
\draw (3.55,1)--(3.9,1) [-latex,red];
\draw (3.55,1.3)--(3.9,1.3) [-latex,red];
\draw (3.55,0.7)--(3.9,0.7) [-latex,red];
\end{tikzpicture}
\caption{An electromagnetic balance.}
\label{figure:0}
\end{figure}

We further assume that the game takes place in an electromagnetic settings. The geniune coins are neutral while the counterfeit is charged either positively or negatively. The target of the human player is to distinguish a (positively or negatively) charged coin from other neutral coins, but the only device available is a test electron which reacts to only the charged coin as Figure. \ref{figure:0}. 

An electromagnetic balance in the vacuum consists of two distant boxes that contains coins and one test electron $+Q$. After coins are put into the two boxes, the test electron shifts to either side or stay at where it was. Unlike in the gravity field, the charged coin is the only reason that causes the shifting of the test electron. One needs not to ensure the numbers of coins in two boxes are the same. At each round, the human player selects two subsets of coins and observes the shifting of the test electron. This assumption is essential to the balance game since this grants the balance to yield an arbitrary result given the human player's configuration. While the human player cannot claim that the balance is violating the physical law on the result of \emph{one single} balancing. For the historic consideration, we still use the diction of \emph{light, weight} in the following discussion. 

One might intuitively assert that the balance game is harder than the balance question for the human player, since the player has to decide the strategy without the gradual exposure of information. One might even question the compactness of the bound yields by the entropy $2n \approx 3^{q}$. However, as what is going to be presented, it turns out that the balance game is no harder. Moreover, given the electromagnetic setting, the bound derived by entropy seems to be tighter than it has been in balance questions. For example, a human play has a must-win strategy in a $(13,3,\text{unknown})$-balance game using a delicate coding method. While there exists a concise algorithm that guarntees the victory of the balance when $2n>3^{q}$ as well. Even if the human player is incapable of adopting a perfect scheme at $2n=3^{q}-1$, he/she can make the balance suffer a lot by deploying a naive randomnized strategy. 

We begin the analysis with an easier case where we know the counterfeit coin is heavier(negatively charged). 
 
\subsection{The $(n,q,\text{heavy})$-balance game}
The $(n,q,\text{heavy})$-balance game involves $n$ visually identical coins, one of which is heavier than others, the player is allowed to use the balance for $q$ rounds. Assuming the balance has known the strategy of the human player, under which circumstances does it have a winning strategy?

To comprehend this game, it is better to adopt a coding framework. The strategy of the human player is embedded into an $n*q$ strategy matrix $S$, while the strategy of the balance is embedded into a $1*q$ mask code $M$, the player observes $M$ and transcripts his/her strategy matrix $S$ into the $n*q$ observation matrix $O$ and tries to infer the counterfeit coin from it. 

Taking the $(n,q,\text{heavy})$-balance game as an instance. The human player chooses $n$ codes, each of length $q$ from an alphabet of $\left\{\text{L},\text{R},\text{O}\right\}$ to form an $n*q$ matrix $S$. If $S_{i,j}=\text{L}$ then the $i$-th coin is put on the left side of the balance at the $j$-th round. If $S_{i,j}=\text{R}$ then the $i$-th coin is put on the right side of the balance at the $j$-th round. If $S_{i,j}=\text{O}$ then the $i$-th coin is not put on the balance at the $j$-th round. 

The balance (or the evil spirit within), knowing the strategy of the human player as $S$, puts forward a mask $M$ of length $q$ using three characters $\left\{\hat{\text{L}},\hat{\text{R}},\hat{\text{D}}\right\}$, if $M_{j}=\hat{\text{L}}$ then the balance says that the left side is heavier at the $j$-th round, $\hat{\text{R}}$ then the right side is heavier and $\hat{\text{D}}$ represents a draw. After observing $M$, the human player translates the effect of the mask code onto $S$ using the following transcription Table. \ref{table:1}:

\begin{table}  
\Large  
\caption{Transcription table for the $(n,q,\text{heavy})$-balance game.}  
\begin{center}  
\begin{tabular}{c|c|c|c}  
\toprule
\  & $\hat{\text{L}}$ & $\hat{\text{R}}$ & $\hat{\text{D}}$\\ 
\midrule  
L & $+$ & $\times$ & $\times$ \\
R & $\times$ & $+$ & $\times$ \\
O & $\times$ & $\times$ & $+$ \\
\bottomrule
\end{tabular}  
\label{table:1}
\end{center}  
\end{table}

Transcripting $S$ using $M$ is simply assigning each entry $O_{i,j}$ the intersection entry between the $C_{i,j}$-th row and the $M_{j}$-th column in Table. \ref{table:1}.
The physical significance behind is that: $O_{i,j}=+$ means the $i$-th coin is possibly heavier according to the $j$-th examination (since the side on which it lies is judged to be heavier according to the $j$-th balancing, or the balance yields a draw while it is kept off the balance), $O_{i,j}=\times$ means that it is firmly not heavier accordingly. 

As a toy illustration, we demonstrate the example of $(4,2,\text{heavy})$-balance game. Entropy says that human can \emph{possibly} win a $(4,2,\text{heavy})$-balance question. What about forcing the human player to present the strategy without knowing the weighing result? Consider:
\begin{equation}
\label{equation:1}
S=
\begin{pmatrix}
\text{L} & \text{L}\\
\text{L} & \text{R}\\
\text{R} & \text{L}\\
\text{R} & \text{R}
\end{pmatrix},
\end{equation}
and
$$M=\left(\hat{\text{L}},\hat{\text{R}} \right).$$
Then the corresponding transcripted matrix is:
$$O=
\begin{pmatrix}
+ & \times\\
+ & +\\
\times & \times\\
\times & +
\end{pmatrix}.$$
Hence the only coin that is possibly heavier is the second one. (One should check this process to examine whether this formulation is consistent with the setting!) In this case the human player wins the game and the balance loses with $M=(\hat{\text{L}},\hat{\text{R}})$. 

Generally, the $i$-th coin is possibly heavier iff the $i$-th row of $O$ contains only $+$. So the player wins if the transcripted matrix contains zero or only one row with no $\times$. If there are no row with only $+$ then the player knows that the balance must have yielded at least one incorrect result (e.g., let $S$ be defined as previous and $M$ be $(\hat{\text{D}},\hat{\text{L}})$). If there are multiple rows that contains only $+$ then the player cannot distinguish between them and the balance wins. In the $(4,2,\text{heavy})$-balance game, adopting the strategy as \eqref{equation:1} guarantees the victory of the human player. One can check this argument by examing all nine possible masks and seeing that the number of $(+\ \ +)$ rows as the function of $M$ returns either one or zero. 

To delve into the game theory aspect, we ask this question: 

\emph{Given $S$, is it possible for the balance to select a mask code such that the player can not locate the counterfeit coin from the transcripted matrix $O$?} 

The case before shows that for some design of $S$, the balance is determined to fail. In other words, once the human player constructs such an $S$, he/she can claim victory without seeing the weighing result or actually conducting deduction. On the other hand, in order to win the $(n,q,\text{heavy})$-balance game, the balance has to design mask code $M(S)$ for each strategy $S$ of the human player such $O(M(S))$ contains more than one pure $+$ rows. 

To study this property, we resort to a probabilistic method. The idea is to assume that the balance encodes a mask randomly (remind that the balance does not has to fix a counterfeit coin beforehand, instead, it only tries to fool the human player without violating logic). We now exert a probability measure on the space of the mask code of the balance 
$$\Omega=\left\{\hat{\text{L}},\hat{\text{R}},\hat{\text{D}} \right\}^{q}.$$
Let each position of $M$ be selected independently and uniformly from $A,B,C$, so each $M\in\Omega$ has probability $3^{-q}$. 

For the $i$-th coin, consider the event $B_{i}$: \emph{ the $i$-th coin is possibly the heavier coin}. 
The probability of such an event is simply $3^{-q}$. Because be $S_{i,j}$ L, R or O, its probability of being considered as possibly heavier is uniformly $\frac{1}{3}$, corresponding to $M_{j}$ be $\hat{\text{L}}, \hat{\text{R}}$ or $\hat{\text{D}}$. In other words, $B_{i}$ is true iff the mask $M$ appears as an image of $S_{i}$ under the one-to-one mapping $\text{L}\rightarrow \hat{\text{L}},\text{R}\rightarrow \hat{\text{R}},\text{O}\rightarrow \hat{\text{D}}$. The indicator random variable of this event is $X_{i}:\Omega\rightarrow\left\{0,1\right\}$, let $X=\sum_{i=1}^{n}X_{i}$ be the random variable that counts the number of coins possibly heavier, now:
$$\mathbb{E}[X_{i}]=\text{Pr}(B_{j})=3^{-q},$$
so
$$\mathbb{E}[X]=\sum_{i=1}^{n}\mathbb{E}[X_{i}]=\frac{n}{3^{q}}.$$
If $\mathbb{E}[X] > 1$ then there must exist a mask $M'\in\Omega$ such that $X(M')>1$, which means that more than one coins are possibly heavier and the human player lacks sufficient information then the balance wins the game. In this naive case (we know that the counterfeit coin is heavier). the bound seems to be the same as the one yields by the entropy. On the other hand, if $n,q$ makes $\mathbb{E}[X]\leq 1$, we only know that it is possible for the balance to lose, and we can only conclude that the balance does not have a must-win strategy, but be the human player unwise, the balance can still choose a $M$ to win the game. Additionally, when $X(M')=0$, the human player can argue that the balance violates logic so it loses. 

Some interesting and straightforward results can be readed from this setting as well: 

\textbf{Theorem:} If $n=2^{q}$ then there exists a must-win strategy for the human player, where at each round all $n$ coins are put onto the balance. 

\textbf{Proof:} Let the $i$-th row of $S$ be the binary representation of $i$ with $0\rightarrow \text{L}$ and $1\rightarrow \text{R}$. Since no coin is ever kept off the balance, the character $\hat{\text{D}}$ has been removed from the balance' s alphabet (or the value of $X$ becomes zero and the balance loses). Mapping $\hat{\text{L}}$($\hat{\text{R}}$) in $M$ into $0$($1$) then there must exist one row in $S$ that is consistent with $M$ and its transcription is $q$ consecutive $+$s, this is the only coin that is possibly heavier since other transcripted row contains at least one zero. \qed

\textbf{Corollary:} If $n\leq 2^{q}$ then there exists a must-win strategy for the human player, where at each round all $n$ chips are put onto the balance. 

\textbf{Proof:} Using binary coding, the rest is the same as the theorem before. \qed

Finally we have:

\textbf{Theorem:} If $n\leq 3^{q}$ then there exists a must-win strategy for the human player. Otherwise the balance has a must-win strategy.

\textbf{Proof:} Using the ternary code of $i$ as $S_{i}$ ($0\rightarrow\text{L},1\rightarrow\text{R},2\rightarrow\text{O}$), since $n\leq 3^{q}$ such a configuration is always legal. Now if $M$ is the ternary code of $l\leq n$ (taking $0\rightarrow\hat{\text{L}}, 1\rightarrow \hat{\text{R}},2\rightarrow\hat{\text{D}}$) then the $l$-th coin is the only candidate according our previous construction. If $l > n$ then $M$ blocks all candidate coins and the human player can claim that the balance is cheating. 

If $n > 3^{q}$ then there must be two indices $i_{1},i_{2}$ such that $S_{i_{1}}=S_{i_{2}}$ componentwise. Now let $M$ be constructed as taking $\text{L}\rightarrow\hat{\text{L}} , \text{R}\rightarrow \hat{\text{R}},\text{O}\rightarrow \hat{\text{D}}$ then the player can neither distinguish the $i_{1}$-th coin from the $i_{2}$-the coin nor argue that the balance is lying.

\qed

The bound $n=3^{q}$ is a tight one in a sense that if $n\leq 3^{q}$ then the human must win, if $n>3^{q}$ then the balance must win. Therefore when both the player and the balance play wisely, the condition $n\leq 3^{q}$ is both necessary and sufficient for the player to win, \emph{vice versa}. This is the same as the original $(n,q,\text{heavy})$-balance question where the player adopts an adaptive strategy in a gravity field. 

\textbf{Remark A:} Although one might eagerly move to the $(n,q,\text{light})$-balance game, we have to point out that such a generalization is not straightforward. In the $(n,q,\text{heavy})$ setting we can always assume that the heavier coin' s weight is larger than the summation of all the rest so no matter how the player put coins on either side of the balance, the balance can always use character $\hat{\text{L}}$ or $\hat{\text{R}}$. But $(n,q,\text{light})$ setting rejects the balance to assign character $\hat{\text{L}}$ to a configuration as putting one coin on the left and two coins on the right. So it is better to adopt the electromagnetic setting.

\textbf{Remark B:} We have been exerting a probability measure on $\Omega$, the space of mask codes. One can also making the space of $S$ into a probability space. The result is exactly the same. However, it is more intuitive to assume that the human player has to give the sequence of coin configurations. While it seems physically improper to ask the human player to device such a sequence given the weighing result of the balance. But we are going to see how a random strategy of the human player can equally torture the balance. 

\subsection{The $(n,q,\text{unknown})$-balance game}
What if the prior information is hiden from the human player such he/she has to infer whether the counterfeit coin is heavier or lighter? Adopting the same way of constructing $S$ as before (since the human player can do no more than putting coins onto two sides of the balance or off the balance), but the transcription table is more complex since not only $+$ and $\times$ are possible characters. We have to include the $-$ character into the observation matrix to represent \emph{possibly lighter}. If the balance yields a draw, we can no longer assign a uniform $+$ or $-$ character to those coins off the balance, instead they are assigned a $\pm$ sign to suggest that they could be heavier or lighter. So the transcription table reads as Table. \ref{table:2}:
\begin{table}[htbp]
\Large  
\caption{Transcription table for the $(n,q,\text{unknown})$-balance game.}  
\begin{center}  
\begin{tabular}{c|c|c|c}  
\toprule
\  & $\hat{\text{L}}$ & $\hat{\text{R}}$ & $\hat{\text{D}}$\\ 
\midrule
L & $+$ & $-$ & $\times$ \\
R & $-$ & $+$ & $\times$ \\
O & $\times$ & $\times$ & $\pm$ \\
\bottomrule 
\end{tabular}  
\end{center}  
\label{table:2}
\end{table}

The transcription rule is exactly the same as the $(n,q,\text{heavy})$-balance game. For example, when $S_{i,j}=\text{R}$ and $M_{j}=\hat{\text{L}}$, then $O_{i,j}$ is assigned the character $-$. $O_{i,j}=+$($-,\pm$) means that the $i$-th coin is possibly heavier (lighter, heavier or lighter) according to the $j$-th examination. While $O_{i,j}=\times$ means that it is firmly not counterfeit accordingly.

We illustrate a toy example $(8,3,\text{unkown})$-balance game to examine the accuracy of this paradigm, let the strategy matrix of the player be:
$$S=
\begin{pmatrix}
\text{R} & \text{L} & \text{L}\\
\text{R} & \text{L} & \text{R}\\
\text{R} & \text{L} & \text{O}\\
\text{R} & \text{R} & \text{L}\\
\text{L} & \text{R} & \text{R}\\
\text{L} & \text{R} & \text{O}\\
\text{L} & \text{O} & \text{L}\\
\text{L} & \text{O} & \text{R}
\end{pmatrix},$$
and let
$$M=\left(\hat{\text{L}},\hat{\text{R}},\hat{\text{L}} \right).$$
Then the transcripted matrix reads:
$$O=
\begin{pmatrix}
- & - & \times\\
- & - & \times\\
- & - & \pm \\
- & + & \times\\
+ & + & \times\\
+ & + & \pm \\
+ & \times & \times\\
+ & \times & \times\\
\end{pmatrix}.$$
Hence it is possible that the  3rd coin is lighter or the 6th coin is heavier. In this case the balance successfully beats the human player and wins the game. The entropy knowledge tells that the human player \emph{might} have a must-win strategy against a balance, which fact is implied by the current framework if the following delicate strategy is selected:
$$S=
\begin{pmatrix}
\text{R} & \text{L} & \text{R} \\
\text{R} & \text{R} & \text{O} \\
\text{R} & \text{O} & \text{L} \\
\text{L} & \text{L} & \text{R} \\
\text{L} & \text{R} & \text{O} \\
\text{L} & \text{O} & \text{L} \\
\text{O} & \text{L} & \text{L} \\
\text{O} & \text{R} & \text{L}\\
\end{pmatrix}.$$
It might be a little anti-intuitive that in the third round, only two coins are put onto the right side of the balance while four coins are on the left side.  One can check (although tedious) that for any of the 27 possible masks of the balance, the number of coins that are possibly heavier/lighter is either zero or one, so either the balance yields a logically wrong balancing sequence  (e.g., $M=\hat{\text{D}}\hat{\text{D}}\hat{\text{L}}$) or the sequence is sufficient for the human player to decide the counterfeit coin (e.g., $M=\hat{\text{D}}\hat{\text{L}}\hat{\text{R}}$). 

We first shift to a statement that reveal the essense of $(n,q,\text{unknown})$-balance game:

\textbf{Theorem: } If $2n \leq 3^{q}-1$ then the human player has a must-win strategy, otherwise the balance has a must-win strategy.

\textbf{Proof: } Consider two ternary codes $S_{1}, S_{2}$ (with alphabet L, R, O) with length $q$. If they satisfy the following conditions, we say that they are \emph{partially complementary to each other}:
\begin{itemize}
\item Whenever $S_{1,j}=\text{O}$, $S_{2,j}$ has to be O, \emph{vice versa}.
\item Whenever $S_{1,j}=\text{L}$(R), $S_{2,j}$ has to be R(L). 
\end{itemize}
Concisely, the non-O part of $S_{1},S_{2}$ are complementary to each other. Now if both $S_{1}$ and $S_{2}$ appear as rows of $S$ then the balance wins the game by adopting the following mask $M$:
$$
M_{j}^{1}=
\left\{
\begin{aligned}
&\hat{\text{L}}, \quad\ \ S_{1,j}=\text{L},S_{2,j}=\text{R},\\
&\hat{\text{R}}, \quad\ \ S_{1,j}=\text{R},S_{2,j}=\text{L},\\
&\hat{\text{D}}, \quad\ \ S_{1,j}=S_{2,j}=\text{O}.\\
\end{aligned}
\right. $$
Then the human player cannot distinguish where the 1st coin is heavier or the 2nd coin is lighter. Thus if two ternary codes of length $q$ with alphabet L, R, O are partially complementary to each other then they cannot appear simultaneously as rows of $S$. That is to say, the size of the set of legal codes for coins is no larger than:
$$\frac{3^{q}-1}{2}+1.$$

If $n\geq \frac{3^{q}-1}{2}+1$ then the  balance can win in either one of the following two ways: 
\begin{itemize}
\item If there exist $i_{1}$ and $i_{2}$ such that $S_{i_{1}}$=$S_{i_{2}}$ or $S_{i_{1}}$ and $S_{i_{2}}$ are partially complementary to each other then the balance wins by transforming $S_{i_{1}}$ into $M$ with $\text{L}\rightarrow\hat{\text{L}},\text{R}\rightarrow\hat{\text{R}},\text{O}\rightarrow \hat{\text{D}}$. Under this setting the $i_{1}$-th coin and the $i_{2}$-th coin remain a mystery for the human player.
\item If $n=\frac{3^{q}-1}{2}+1$ and there are neither duplication nor partial complementary then there must exists $i$ such that $S_{i}=(\text{O},\text{O}\cdots,\text{O})$ Let $M=(\hat{\text{D}},\hat{\text{D}},\cdots,\hat{\text{D}})$ then the player can only know that the $i$-th coin is different in weight, but whether or not it is heavier remains unknown.
\end{itemize} 

On the other hand, if $n\leq \frac{3^{q}-1}{2}$ then the human player has a must-win strategy. He/she only has to select $n$ codes from the codebook $\left\{\text{L},\text{R},\text{O} \right\}^{q}/(\text{O},\text{O},\cdots,\text{O})$ module partial complementarity. Now if the mask code $M$ is partially complementary/identical to some $S_{i}$ then the unique $i$-th coin is lighter/heavier. Otherwise the player can conclude that the balance is lying on some experiments. \qed

We illustrate an example of $(13,3,\text{unknown})$-balance game with the configuration designed in the proof of the previous theorem:
$$S=
\begin{pmatrix}
\text{L} & \text{L} & \text{L} \\
\text{L} & \text{L} & \text{R} \\
\text{L} & \text{R} & \text{L} \\
\text{L} & \text{R} & \text{R} \\
\text{O} & \text{R} & \text{R} \\
\text{O} & \text{L} & \text{R} \\
\text{R} & \text{O} & \text{L} \\
\text{L} & \text{O} & \text{L} \\
\text{R} & \text{L} & \text{O} \\
\text{L} & \text{L} & \text{O} \\
\text{O} & \text{O} & \text{R} \\
\text{L} & \text{O} & \text{O} \\
\text{O} & \text{R} & \text{O} \\
\end{pmatrix},
$$
Then a mask, e.g., $M=(\hat{\text{D}},\hat{\text{L}},\hat{\text{R}})$ can identify only $(\text{O},\text{R},\text{L})$ or $(\text{O},\text{L},\text{R})$, in which one and only one code must have appeared as one row of $S$. 

\textbf{Corollary: } If $n=\frac{3^{q}-1}{2}$ then the number of different strategy matrices is:
$$2^{n}\cdot n!q!.$$

\textbf{Proof: } Let $l$ be the number of $\left\{\text{L},\text{R}\right\}$ in one row then the subspace with $(q-l)$ Os has size $\binom{q}{l}\cdot 2^{l}$. In such a space exist $\binom{q}{l}\cdot 2^{l-1}$ pairs of complementary codes from which one have to select one from each pair, hence there are altogether $2^{\binom{q}{l}\cdot 2^{l-1}}$ different choices. Multipling over different $l$ yields that the number of strategies is:
$$\prod_{l=1}^{q}2^{\binom{q}{l}\cdot 2^{l-1}}.$$
The binomial theorem reduces this value to $2^{n}$. Finally, one strategy can be compile into $2^{n}\cdot n!q!$ matrices.
  \qed

\begin{figure}[htb]
\centering
\includegraphics[width=0.45\textwidth] {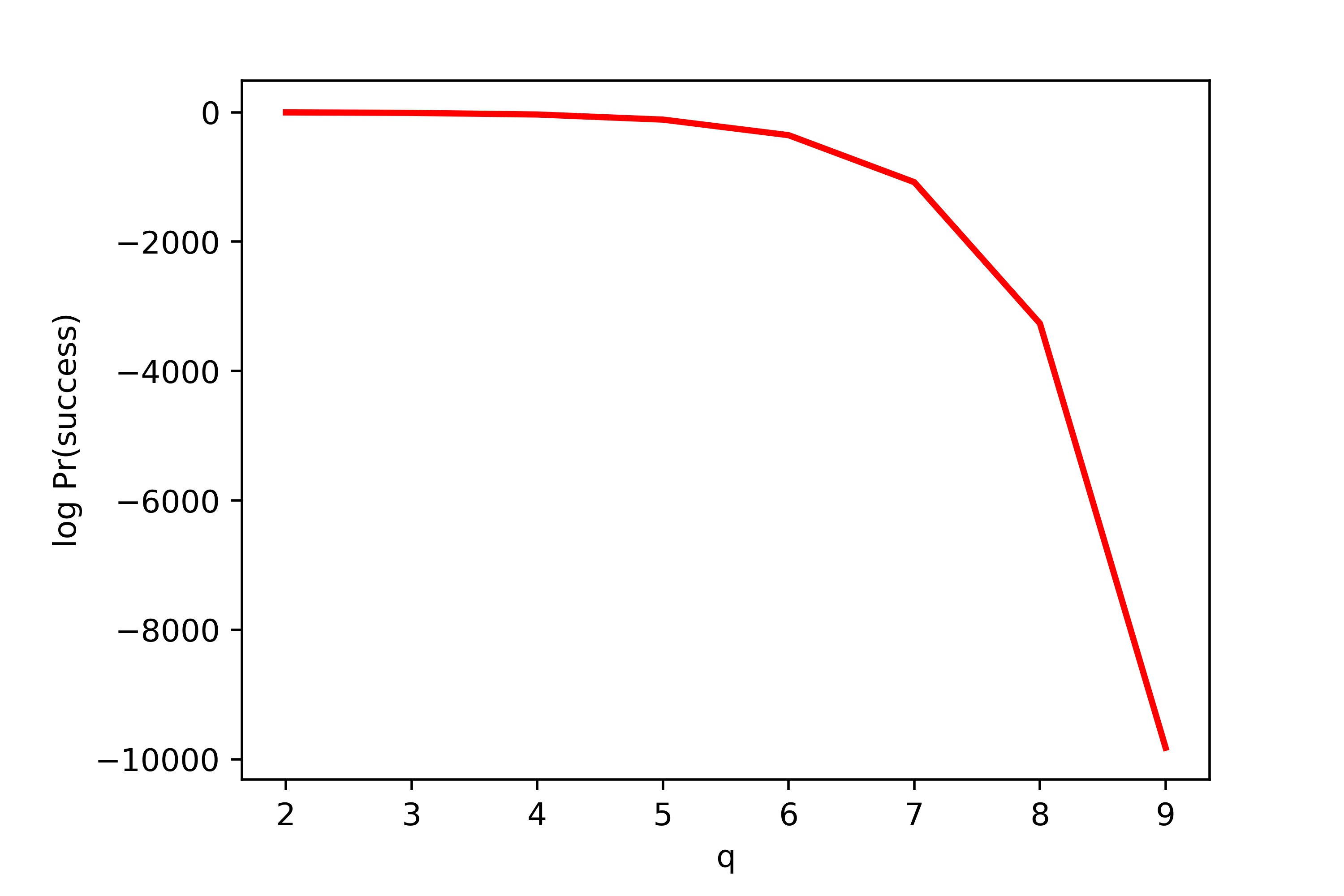}
\caption{$\log \text{Pr(success)}$.} 
\label{figure:ratio}
\end{figure}

\textbf{Remark:} The ratio between the number of perfect strategy and that of the entire strategy space, or the probability that a complete random strategy is optimal is:
$$\text{Pr}(\text{success})=\frac{2^{n}n!q!}{3^{n\cdot q}}.$$
It is better to evaluate its logarithm which is a function of $q$. The result is in Figure. \ref{figure:ratio}.

Generally, it is almost impossible for a random strategy appear to be optimal given a large $q$. However, it is still inspiring to study the randomnized strategy. 

\subsection{Randomnized strategy}
We now take a probabilistic perspective into this game. What makes the $(n,q,\text{unknown})$-balance game harder to analyze with probabilistic method is that the transcription table is not rowwise homogeneous (i.e., different rows might preserve different sets of elements), hence the probability that $O_{i,j}$ be assigned each one of the four character $\left\{\times,+,-,\pm \right\}$ is no longer independent of $S_{i,j}$. 

Fortunately, the first two rows of the transcription Tables \ref{table:2} contain the same set of characters, so we let $q_{i}$ be the times that the $i$-th coin be put onto the balance (note that $q_{i}$ is a statistics of $S$, when we compute conditioning on $q_{i}$ we actually compute conditioning on $S$, hence the factor $\binom{q}{q_{i}}$ is not needed). For the balance, at each round, the character $\hat{\text{L}}$ and $\hat{\text{R}}$ is inserted into $M$ with probabiltiy $p$ and $C$ with probability $(1-2p)$. Therefore the probability that the $i$-th coin turns out to be possibly heavier/lighter is
$$\left\{
\begin{aligned}
&2(1-2p)^{q-q_{i}}p^{q_{i}},\quad 0<q_{i}\leq q, \\
&0,\quad\quad\quad\quad\quad\quad\quad\quad q_{i}=0.
\end{aligned}\right.
$$

To see why it holds, note that the $i$-th coin is considered to be heavier/lighter iff the $i$-th row of $O$ contains only $\left\{+,\pm \right\}$ or $\left\{-,\pm \right\}$. That is to say, it must get $q_{i}$ consecutive $+$/$-$ signs when on the balance, and the balance has to yield a draw for all $q-q_{i}$ times whenever it is off the balance. If $q_{i}=0$ then there is no way of determing whether it is heavier/lighter even if all the rest $(n-1)$ coins weigh equally.

Having this observation, we are ready to analyze and optimize a random strategy. 

Imagine a less intellectual human player who hates delicate arrangement. Instead, the player can only conduct a random strategy in which at each round, each coin is put onto the left side of the balance with probability $\frac{r}{2}$, onto the right side with probability $\frac{r}{2}$ and off the balance with probability $(1-r)$. We now conduct an analysis over this player. Let $\textbf{q}=\left\{q_{i} \right\}_{i=1}^{n}$ be the sufficient statistics from $S$, whose possible assignments is a subset of $\left\{0,1,2,\cdots,q\right\}^{n}$ chosen by the player. 

\textbf{Theorem:} Using the previous notations, the balance can always win the $(n,q,\text{unknown})$-balance game where the strategy of the human player has to be predetermined if:
\begin{equation}
\label{equation:2}
\min_{\textbf{q}}\left\{\max_{p}\left\{\sum_{i=1}^{n}2(1-2p)^{q-q_{i}}p^{q_{i}} \right\} \right\} >1.
\end{equation}

\textbf{Proof:} Given $\textbf{q}$ and $p$, the probability that the $i$-th coin is possibly heavier/lighter is $2(1-2p)^{q-q_{i}}p^{q_{i}}$. Then the expectation of the number of possibly heavier/lighter coins is just 
$$f(\textbf{q},p)=\sum_{i=1}^{n}2(1-2p)^{q-q_{i}}p^{q_{i}}.$$ 
If for any configuration of $\textbf{q}$, there exists one $p$ such $f(\textbf{q},p)>1$ then it is always possible to choose one mask code such that more than one coins are dubious and are indistinguishable for the human player. Hence the balance wins the game.\qed 

Once the coding delicacy in $S$ breaks down, we can do no more than addressing this sufficient condition for the balance to win. However, it is possible for the human player to modify $r$ and hence $\textbf{q}$ to increase the threshold $n$ for \eqref{equation:2} given $q$. After all, \eqref{equation:2} is quite intimidating so we can only try to derive some asymptotical behavior. 

Begin with assuming $r\approx\frac{1}{2}$, so $q_{i}\approx \frac{q}{2}$, i.e., each coin is put onto the balance for approximately half of the rounds, then
$$f(\textbf{q},p)=2n(p-2p^{2})^{\frac{q}{2}}\leq 2n 8^{-\frac{q}{2}}.$$
Straightforward algebra yields that if 
$$2n>(2\sqrt{2})^{q},$$
then the balance has a must-win strategy. 

One should note that this bound is miserably worse than the one derived by the section before ($2n>3^{q}$). So we can conclude that if the less intelligent human player chooses a bad random strategy (specifically a bad $r$) then the game becomes much easier for the balance in multiple perspectives:
\begin{enumerate}
\item The balance can possibly win even if $2n<3^{q}$ (where a clever human player can always win) since a random player might unfortunately include duplicated/partial complementary rows of $S$.
\item The must-win threshold for the balance could be exponentially declined, hence the region of $n$ where the balance has a must-win strategy sharply increases. 
\end{enumerate}

We now optimize \eqref{equation:2} w.r.t. $r$ to cope with the second privileged aspect of the balance so the random player behaves less lamantable. Plugging $rq$ into $q_{i}$, then:
$$f(q)\leq 2n\left[(1-r)\left(\frac{r}{2(1-r)} \right)^{r} \right]^{q}.$$

Hence the sufficient condition for the balance to win reduces to:
\begin{equation}
\label{equation:3}
2n > \left[(1-r)\left(\frac{r}{2(1-r)} \right)^{r} \right]^{-q}=g(r)^{q},
\end{equation}
where $g(r)=(1-r)^{-1}\left(\frac{r}{2(1-r)} \right)^{-r}$.

\begin{figure}[htb]
\centering
\includegraphics[width=0.5\textwidth] {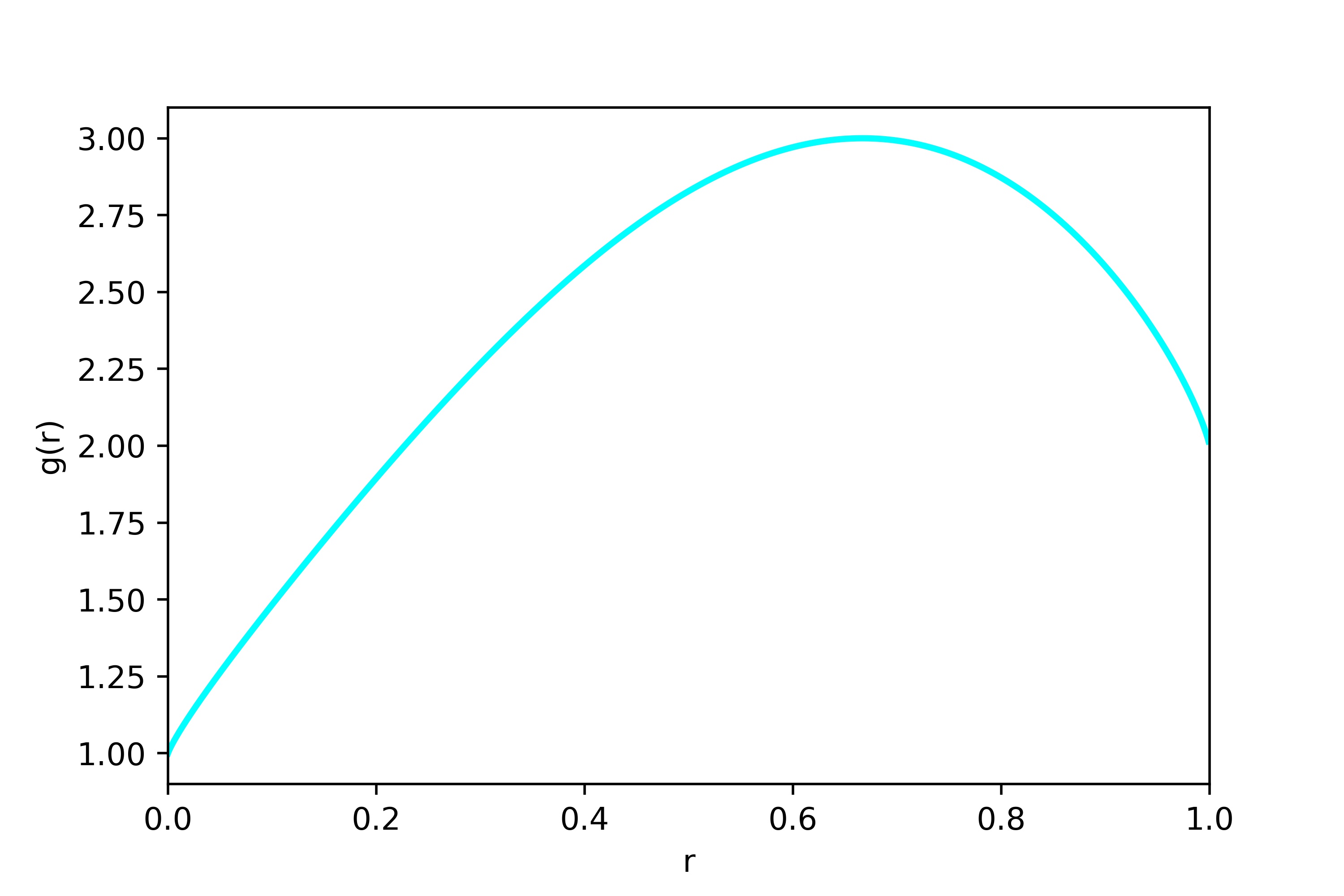}
\caption{$g(r)$.} 
\label{figure:1}
\end{figure}

By plotting $g(r)$ explicitly in figure. \ref{figure:1}, we can conclude that the bound $2n > 3^{q}$ only appear at the peak value of $g$ since $\max_{r}\left\{g(r) \right\}=3$, in other area, the bound can be significantly lower, hence the balance is much easier to win than one might imagine (once $r$ is removed from $\frac{1}{3}$ for a non-trivial distance, the must-win threshold for the balance declines exponentially). Another particularly interesting region of $r$ besides $\frac{1}{2}$ and $\arg\max_{r}\left\{g(r) \right\}$ is $r\rightarrow 1$, in which case $g(r)\rightarrow 2$. Then $p\rightarrow \frac{1}{2}$ (since almost no coin is put out of the balance, the probability that the balance select $\hat{\text{D}}$ into the mask code approaches zero), hence given $2n > 2^{q}$ ensures the victory of the balance. In conclusion, it is unwise to eagerly put all chips onto the balance at all time, the best ratio between $q_{i}$ and $q$ should be approximately $\frac{2}{3}$, since $\frac{\text{d}g}{\text{d}r}$ is zero at $r=\frac{2}{3}$. 

\textbf{Remark:} The value $r=\frac{2}{3}$ sheds light on an \emph{optimal random strategy}. The optimal random strategy is to put each coin on either side of the balance or off the balance with probability $\frac{1}{3}$ at all rounds. Only in this way the must-win strategy can be improved to the level of the theoretical optimal one. Otherwise the balance can easily invade the region ($2n<3^{q}$, given $q$) and marks extra $n$s as ''must-win area for the balance''. What might be an interesting result is that the winning bound for the balance yielded by a total random player is the same as the strict one yield by the section before, but the meaning of this bound is different. For a rational player, $2n<3^{q}$ means a victory, while it only means that the balance cannot necessarily win for a random player. 

Although setting $r=\frac{2}{3}$ cannot ensure that $q_{i}$ converges to $\frac{2}{3}q$ with probability one, we have that when $q\rightarrow\infty$, it is always sure that
$$\frac{q_{i}}{q}=\frac{2}{3}.$$
To see this point, for $s=1,2,\cdots,q$, let $Z_{s}$ be the random variable that takes value one with probability $\frac{2}{3}$ and value zero with probability $\frac{1}{3}$. Let $Z'_{s}=Z_{s}-\frac{2}{3}$. Now let
$$Z=\sum_{s=1}^{q}Z_{s},$$
$$Z'=\sum_{s=1}^{q}Z'_{s},$$
according to the Chernoff bounding theory:
$$\text{Pr}(|Z'|>\epsilon)<2\text{e}^{-\frac{2\epsilon^{2}}{q}}.$$
Finally, let $\epsilon=\delta\cdot q$, where $\delta$ is an arbitrary small number:
$$\text{Pr}(|\frac{Z}{q}-\frac{2}{3}|>\delta)<2\text{e}^{-2\delta^{2}q}.$$
Let $q$ approaching infinity (naturally $n$ approaching infinity faster accordingly) yields the observation on stability.

\section{The Dishonest Balance Game}
\label{section:4}
We now proceed to the dishonest balance game that further reveal the power of probability based analysis. To make the start painless, we begin with $(n,q,\text{heavy})$-balance game, the balance is granted $k$ times of telling lies, this makes up the $(n,q,k,\text{heavy})$-balance game.

\subsection{The $(n,q,k,\text{heavy})$-balance game}
\textbf{Theorem: } If 
$$n\sum_{j=0}^{k}\binom{q}{j}\geq 3^{q},$$
then the balance has a must-win strategy.

\textbf{Proof: } Exerting a probability measure on $\Omega$ as before, each $M\in\Omega$ is assigned probability $3^{-q}$. Let $D_{i}$ be the event \emph{the $i$-th coin is possibly the heavier counterfeit coin}. Since the balance can tell at most $k$ lies, the $i$-th coin is possibly heavier iff the $i$-row in $O$ contains no more than $k$ $\times$s. Now the probability of $D_{i}$ is:
$$\text{Pr}(D_{i})=\sum_{j=0}^{k}\binom{q}{j}3^{-q},$$
where $j$ counts the number of $\times$s in $O_{i}$. Let $Y_{i}$ be the indicator random variable of $D_{i}$ and $Y=\sum_{i=1}^{n}Y_{i}$. Now the random variable $Y$ that counts the number of coins possibly heavier has expectation:
$$\mathbb{E}[Y]=\sum_{i=1}^{n}\text{Pr}(D_{i})=n\sum_{j=0}^{k}\binom{q}{j}3^{-q}.$$
If $\mathbb{E}[Y] >1$ then there exists one mask $M''\in\Omega$ such $Y(M'')>1$ hence the balance is determined to win. \qed
 
\textbf{Corollary: } If the player has to put all $n$ coins onto the balance during the $(n,q,k,\text{heavy})$-balance game then if
$$n\sum_{j=0}^{k}\binom{q}{j}\geq 2^{q},$$
then the balance has a must-win strategy.

\begin{table}  
\Large  
\caption{Transcription table for a special case of the $(n,q,k,\text{heavy})$-balance game}  
\begin{center}  
\begin{tabular}{c|c|c}  
\toprule
\  & $\hat{\text{L}}$ & $\hat{\text{R}}$ \\ 
\midrule
\text{L} & $+$ & $\times$  \\
\text{R} & $\times$ & $+$  \\
\bottomrule
\end{tabular}  
\end{center}
\label{table:3}  
\end{table}

\textbf{Proof: } In this case the balance erases $\hat{\text{D}}$ from the alphabet and adopting the transcription Table. \ref{table:3}:

Exerting a probability measure on $\Omega=\left\{\hat{\text{L}},\hat{\text{R}}\right\}$ with each characters being selected independently and uniformly. Let $D_{i}$ be defined as in the proof of the previous theorem, then
$$\text{Pr}(D_{i})=\sum_{j=0}^{k}\binom{q}{j}2^{-q}.$$
The rest is the same.\qed

\subsection{A note on the liar game}
The motivation and the solution of the dishonest balance game is similar to that of a \emph{liar game}. Let $N=\left\{1,2,\cdots,n\right\}$. In a liar game, Bob selects one specific $a\in N$, Alice tries to figure out $a$ by asking a series of questions ''is $a$ in $Q$?'' where $Q$ is a subset of $N$. Alice can ask altogether $q$ questions in which Bob can tell $k$ lies. On can easily prove the equivalence between the $(n,q,k,\text{heavy})$-balance game and the $(n,q,k)$-liar game, thus the sufficient condition for Bob to win is:
$$n\sum_{i=0}^{k}\binom{q}{i}\geq 2^{q}.$$

The derivation of this bound is identical to the paradigm we adopted in the previous section. Alice gives an $n*q$ matrix $S$, then Bob gives a mask $M$ and the transcription matrix reads Table. \ref{table:4}:

\begin{table}  
\Large  
\caption{Transcription table for the liar game}  
\begin{center}  
\begin{tabular}{c|c|c}  
\toprule
\  & $A$ & $B$ \\ 
\midrule
0 & $0$ & $1$  \\
1 & $1$ & $0$  \\
\bottomrule
\end{tabular}  
\end{center}
\label{table:4}  
\end{table}

Semantically, $S_{i,j}=0$ iff Alice selects $i$ into $Q$ at the $j$-the round, $M_{j}=A$ means that Bob gives a positive answer. The $O_{i,j}$ entry of the transcripted matrix is zero iff $i$ is considered to be possibly $a$ during the $j$-th round of game. Now since Bob can lie for $k$ times, the event that $i$ is possibly $a$ is true if the $i$-th row of the transcripted matrix contains at most $k$ ones. Define the probability space by having Bob choose each position of the mask with $A$ or $B$ independently and uniformly. So the probability of the event that $i$ is possibly $a$ is true is:
$$\sum_{i=0}^{k}\binom{q}{i}2^{-q}.$$
Hence the expectation of the number of elements possibly $a$ is
$$n\sum_{i=0}^{k}\binom{q}{i}2^{-q}.$$
If it is larger than unity then there exists one mask such that more than one numbers are possibly $a$ and Alice cannot distinguish them. 

There is another game, \emph{liar chip game}, which is isomorphic to the liar game and hence the dishonest balance game \cite{alon2004probabilistic}. 

\subsection{An observation between the dishonest balance game and Shannon' s theorem}
Though implicitly, the liar game/liar chip game/dishonest balance game is essentially related to Shannon' s theorem. The identical topic behind both reasoning is essentially communication under the disturbance of noise. Recall that Shannon' s theorem roughly states that in a noisy channel (with $n$ signals and code length of $q$, while $r$ is the bitwise error probability), it is almost always possible to adopt a coding scheme so that every Hamming ball with radius $rq$ in the code space does not interset, hence the reliable communication is ensured and the rate of transmission is higher than $1-H(r)$. 

In the dishonest balance game where all $n$ coins are put onto the balance at all rounds, we can adopt a similar reasoning and arrive at the same bound without mentioning a probability measure over the space of mask code. In $\left\{0,1\right\}^{q}$, a Hamming ball with radius $k$ contains codes at most
$$\sum_{i=0}^{k}\binom{q}{i}.$$
The size of the code space is exactly $2^{q}$, and there are $n$ signals to be encoded. Now if
$$n\sum_{i=0}^{k}\binom{q}{i}>2^{q},$$
then there exist two Hamming balls whose intersection is not empty. Pick one code $M$ from this intersected area and denote the centers of these two balls as $S_{1}$ and $S_{2}$. So the Hamming distance between $M$ and $S_{1}$ and that between $M$ and $S_{2}$ are both no larger than $k$. Following the construction before, if $S_{1}$ and $S_{2}$ are two rows of the player's $S$ and the balance selects the mask code $M$ then the human player cannot distinguish between these two coins and the balance wins. 

In fact, one can show that:
$$\sum_{i=0}^{pn}\binom{n}{i}\sim 2^{n(H(p)+o(1))},$$
thus the bound of the dishonest balance game is almost the same as the bound given by Shannon's Theorem II with error probability $\frac{k}{n}$.

\subsection{The $(n,q,k,\text{unknown})$-dishonest balance game}
Finally, we study the setting where the balance can cheat for $k$ times out of altogether $q$ rounds of balancing, and the player has no extra prior information. We denote this game as $(n,q,k,\text{unknown})$-balance game. When cheating comes into the stage, we can hardly rely on a deterministic coding scheme, so we resort to the probabilistic method for a bound of $n$ for the balance to have a must-win strategy. Analogously, we have the following theorem: 

\textbf{Theorem:} Let $\textbf{q}=(q_{1},\cdots,q_{n})$ be defined as the previous section, define:
$$
h(q_{m},p)=2(1-2p)^{q-q_{m}}p^{q_{m}}\sum_{i=0}^{q_{m}-(k+1)}\binom{q_{m}}{i}.$$
Now if:
$$\min_{\textbf{q}}\left\{\max_{p}\left\{\sum_{m=1}^{n}h(q_{m},p)\right\}\right\} > 1,$$
then the balance has a must-win strategy given the policy of the human player.

\textbf{Proof:} As defined, $q_{m}$ is the number of times that the $m$-th coin be put onto the balance. Define a probability measure on the space of the balance's mask code, where it selects one side to be heavier/lighter with probability $p$. Let $D_{q_{m},p}$ be the event that \emph{the $m$-th coin is considered to be possibly heavier/lighter after the game}. 

Now if $q_{m}\leq k$ then $D_{q_{m},p}$ naturally fails. The $m$-th coin is deduced to be heavier iff the number of $\left\{\times,- \right\}$ in $O_{m}$ (the $m$-th row of $O$) is less than or equal to $k$, and the number of $\left\{+\right\}$ in that row is strictly larger than $k$. For the $m$-th coin to be considered lighter, the situation is similar. 

Hence $D_{q_{m},p}$ is true if the mask takes character $\hat{\text{D}}$ at all positions where $O_{m}$ is $2$ and selects no more than $q_{m}-(k+1)$ positions to transcript the corresponding component of $O_{m}$ into those signs against the dominance decision (the reverse statement is not strictly true, since the components on where the $m$-th coin is not selected can also appear to be false). So
$$\text{Pr}(D_{q_{m},p}) \geq h(q_{m},p).$$
Finally, the random variable that counts the number of possibly heavier/lighter coins has expectation no less than
$$\sum_{m=1}^{n}h(q_{m},p).$$
If $\forall \textbf{q}$ there exists a $p$ such that $\sum_{m=1}^{n}h(q_{m},p) > 1$ then in all cases the balance is determined to win. \qed

\begin{figure}[htb]
\centering
\includegraphics[width=0.5\textwidth] {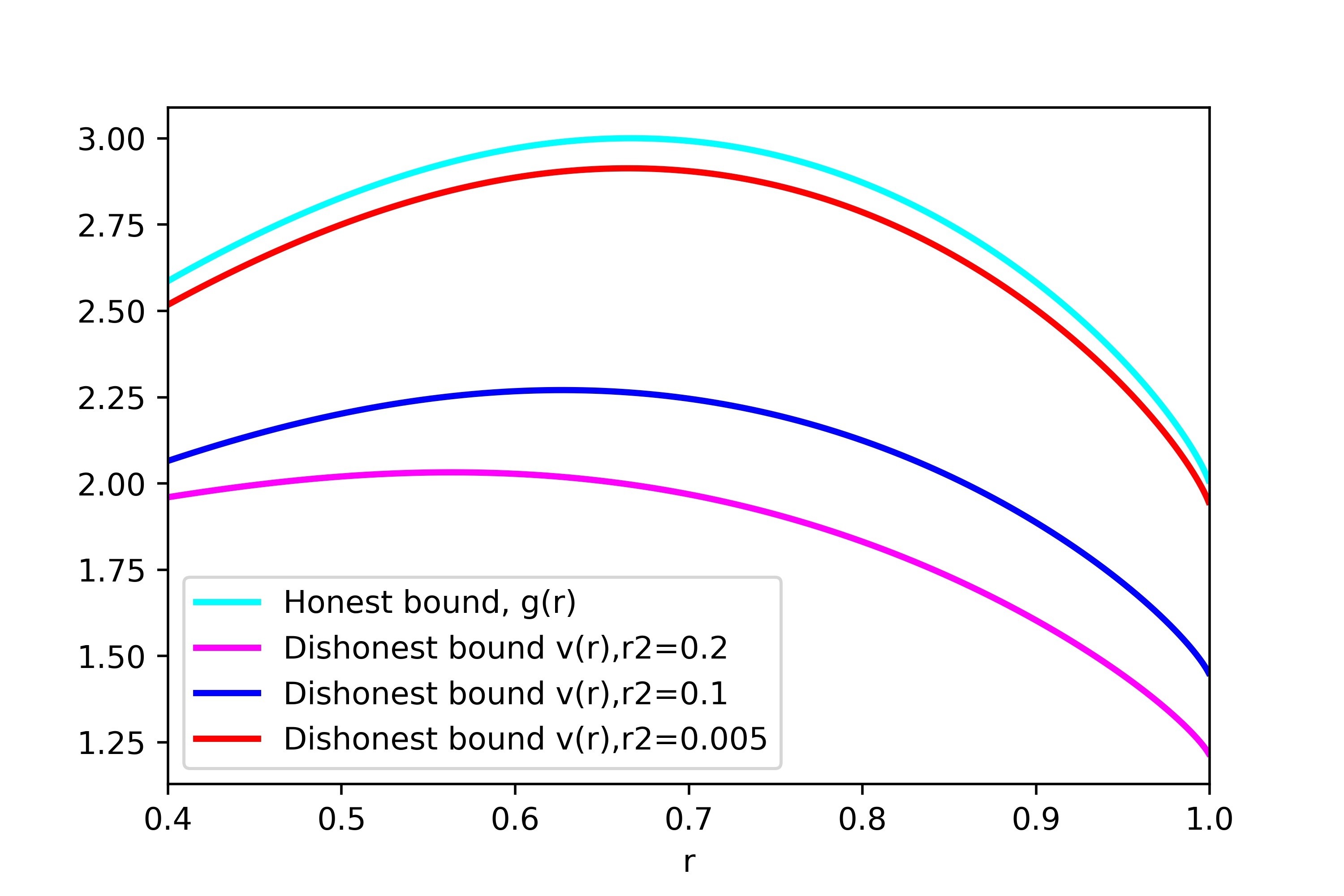}
\caption{$g(r)$ vs. $v(r)$, $r_{2}=0.2,0.1,0.005$.}  
\label{figure:2}
\end{figure}

Finally, let the human player takes a random strategy, we are now interest in the asymptotic performance of the sufficient condition given by the theorem before. Again, let:
$$q_{m}\approx r\cdot q,$$
$$k =r_{2}\cdot q.$$
Of course we should have $r > r_{2}$. Approximating the term $\sum_{i=0}^{q_{m}-(k+1)}\binom{q_{m}}{i}$ by $2^{rqH(\frac{r-r_{2}}{r})}$ and maximizing $h(r,r_{2},p)$ w.r.t. $p$ yields $p=\frac{r}{2}$ as in the previous conclusion on honest $(n,q,\text{unknown})$-balance game. Finally, the only thing that a random player can do to decrease the must-win region of $n$ (given $r_{2}$ and $q$) for the balance is to choose $r$ in order to maximize:
\begin{equation}
\label{equation:4}
\left[(1-r)\left(\frac{r}{2(1-r)} \right)^{r} \right]^{-q}\cdot 2^{-rqH(\frac{r-r_{2}}{r})}=v(r)^{q}.
\end{equation}
One should note that \eqref{equation:4} is a graceful generalization of \eqref{equation:3} (taking $r_{2}=0$ so $k=0$ then \eqref{equation:4} reduces to \eqref{equation:3}). However, it is hard to find an analytic solution of the optimum of \eqref{equation:4}. Instead we plot the value of $v(r)$ in \eqref{equation:4} with $r_{2}=0.2,0.1,0.005$ in Figure. \ref{figure:2}, as an intuitive illustration.

One can conlude from Figure. \ref{figure:2} that after adding the pertubation as a result of dishonest, the optimal $r$ is no longer uniformly $\frac{2}{3}$ (e.g., when $r_{2}=0.2$). While Figure. \ref{figure:2} also shows that the decline of $r_{2}$ reduces $v(r)$ to $g(r)$. 

\begin{figure}[htbp]
\centering
\subfigure[$r(r_{2})$.]{
\begin{minipage}[c]{0.4\textwidth}
\centering
\includegraphics[width=6.8cm]{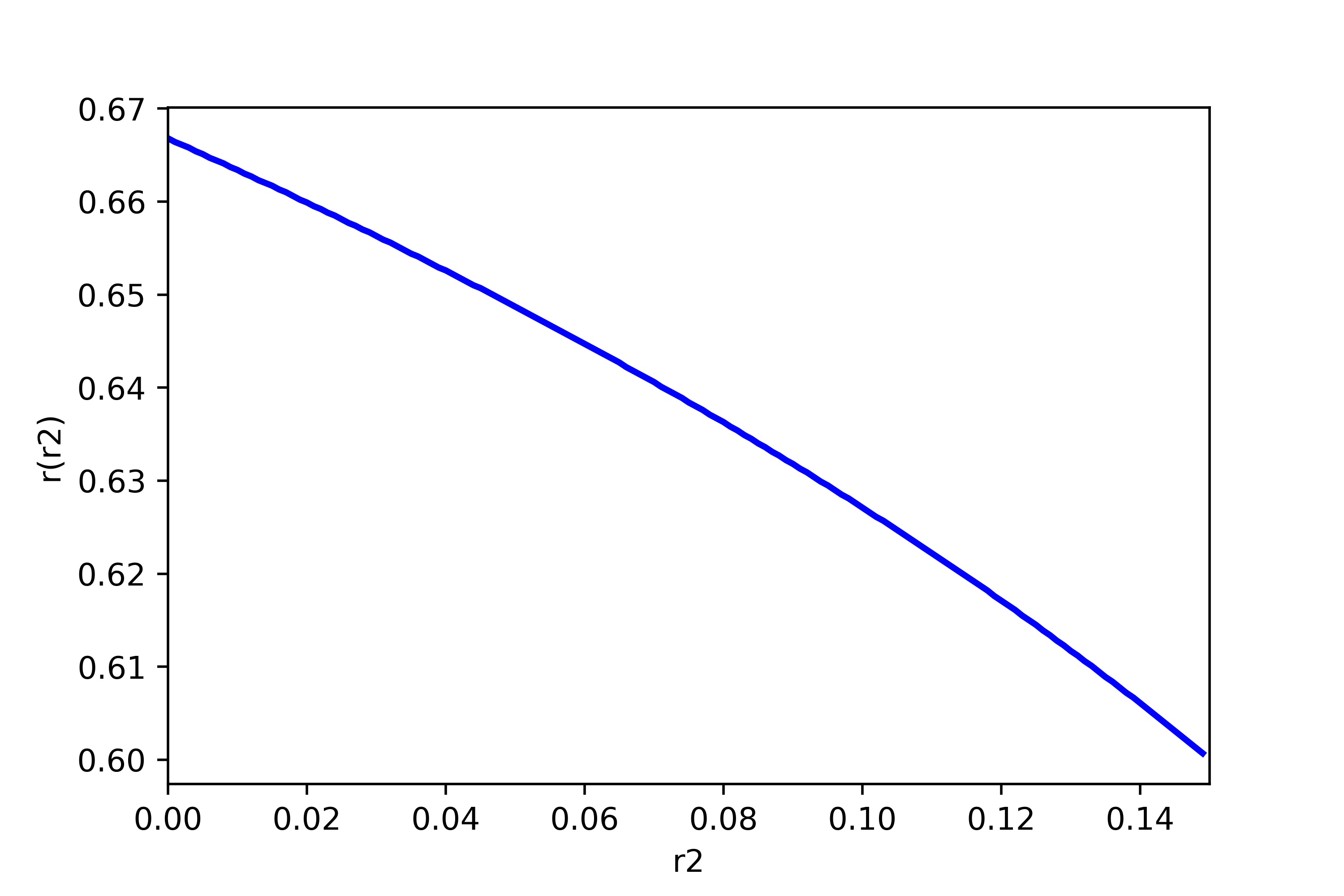}
\end{minipage}
\label{figure:3a}
}
\subfigure[$v(r(r_{2}))$.]{
\begin{minipage}[c]{0.4\textwidth}
\centering
\includegraphics[width=6.8cm]{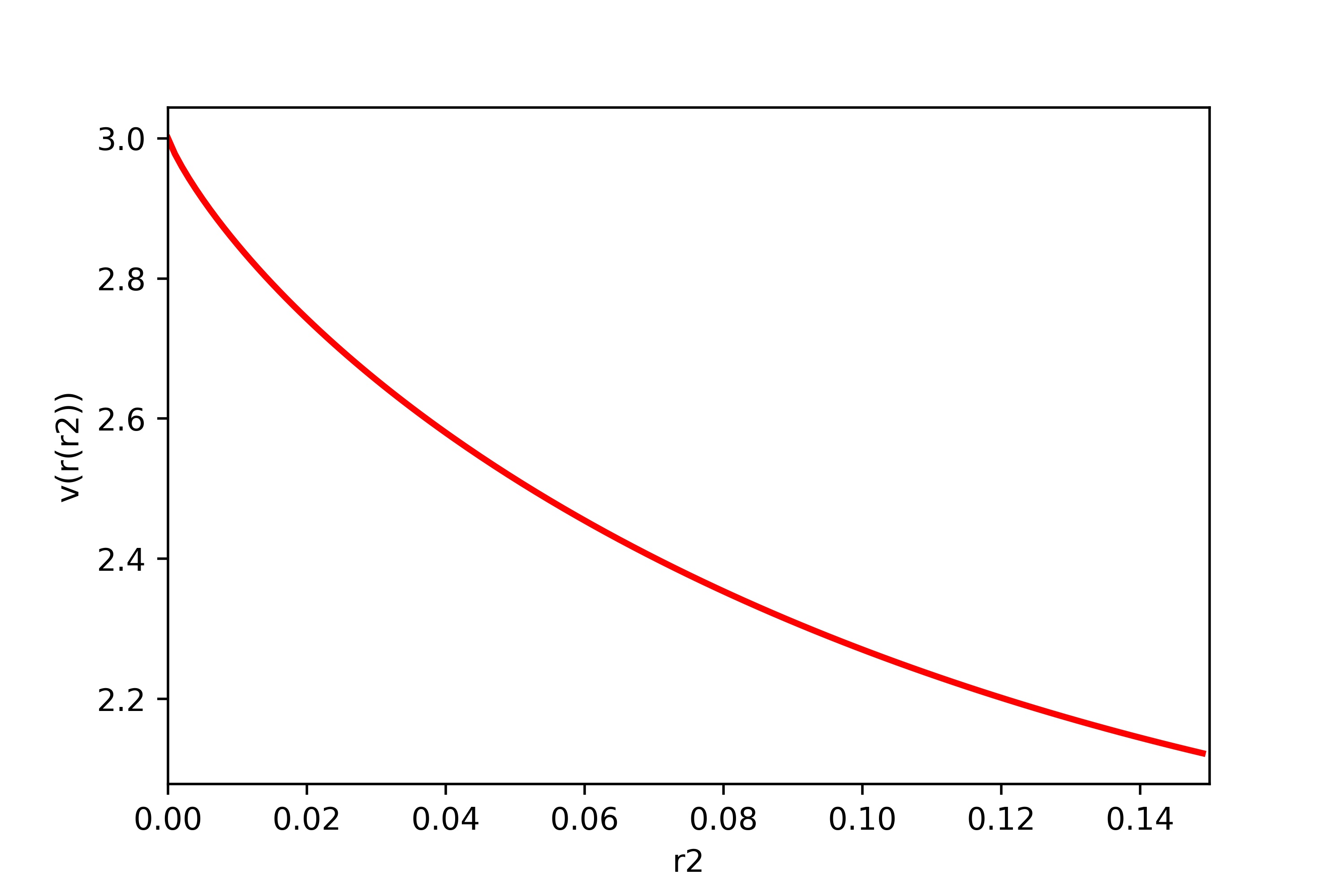}
\end{minipage}
\label{figure:3b}
}
\caption{The variation of the random player's optimal strategy w.r.t. $r_{2}$.}
\label{figure:3}
\end{figure}

Assume that the random human player always plays optimally w.r.t. $r_{2}$ so that:
$$r(r_{2})=\arg\max\left\{v(r) \right\}.$$
We plot $r(r_{2})$ and $v(r(r_{2}))$ in Figure. \ref{figure:3}. 

Generally speaking, it is suggested that with the growth of $r_{2}$, the human player should better adopt a less $r$. That is to say, the more deceptive the balance is, the less should the human put coins onto the balance. The rate of decreasing is visualized in Figure. \ref{figure:3a}. 

Meanwhile, with the dishonest mechanism, it is easier for the balance to win the balance game. The threshold for the balance to have a must-win strategy declines from approximately $\frac{3^{q}}{2}$ to $\frac{a^{q}}{2}$ where $a=\max_{r}\left\{v(r) \right\}$ is a function of $r_{2}$ and is uniformly smaller than three according to Figure. \ref{figure:3b}. 

\section{Conclusion and Discussion}
\label{section:5}
In this paper we study the balance game. By exerting the extra electromagnetic assumption, the winning conditions for the human player and the balance are derived under various settings. An analysis on the random strategy shows that if the human player plays randomly, he/she can still hinder the balance by choosing good parameters. The dishonest balance game, in which the balance can cheat the player, is then studied. The behavior of an randomnized human player in this game is demonstrated to be correlated to the bound in noisy-channel communication. 

Amongst the infinity of extra assumptions/generalizations that can be exerted onto the naive balance-coin setting, it is hard to conclude which combinations of assumptions can give rise to fruitful results without actually delving into them and applying appropriate mathematical tools. 

\subsection{Generalization case: more counterfeit coins?}

For example, let us deviate from the previous discussion and consider the generalization of two counterfeit coins, one heavier, one lighter, and the sum of the weight is the weight of two normal coins. In our setting we can have two counterfeit coins charged $+Q$ and $-Q$ respectively. Is it still safe to adopt the previous framework? One easily see that the transcription table has to become Table. \ref{table:5}.
\begin{table}[htbp]
\Large  
\caption{Transcription table for the $(n,q,\text{unknown})$-balance game.}  
\begin{center}  
\begin{tabular}{c|c|c|c}  
\toprule
\  & $\hat{\text{L}}$ & $\hat{\text{R}}$ & $\hat{\text{D}}$\\ 
\midrule
L & $+$ & $-$ & $\pm$ \\
R & $-$ & $+$ & $\pm$ \\
O & $\pm$ & $\pm$ & $\pm$ \\
\bottomrule 
\end{tabular}  
\end{center}  
\label{table:5}
\end{table}

For the $i$-th coin which has been put onto the balance for $q_{i}$ times, the probability that it be considered heavier is:
$$
\begin{aligned}
&\sum_{j=1}^{q_{i}}\binom{q_{i}}{j}p^{j}(1-2p)^{q_{i}-j}\\
&=(1-2p)^{q_{i}}\cdot\sum_{j=1}^{q_{i}}\binom{q_{i}}{j}\left(\frac{p}{1-2p} \right)^{j}\\
&=(1-2p)^{q_{i}}\left[\left(\frac{1-p}{1-2p}\right)^{q_{i}}-1 \right]\\
&=(1-p)^{q_{i}}-(1-2p)^{q_{i}}\\
&\approx (1-p)^{rq}-(1-2p)^{rq}=\phi(p,r).
\end{aligned}
$$

The problem is: the sufficient condition for the balance to win can no longer be calculated using expectation. The orthodox probabilistic treatment at this stage is to let $X_{i}^{+}$ be the indicator random variable of the event that the $i$-th coin is considered heavier and approximately compute:
\begin{equation}
\label{equation:5}
\begin{aligned}
\text{Pr}(X^{+}&=\sum_{i=0}^{n}X_{i}^{+}=0)\\
&\approx \prod_{i=1}^{n}\text{Pr}(X_{i}^{+}=0)\\
&= (1-\phi(p,r))^{n}.
\end{aligned}
\end{equation}
\begin{equation}
\label{equation:6}
\begin{aligned}
\text{Pr}(X^{+}&=\sum_{i=0}^{n}X_{i}^{+}=1)\\
&\approx \sum_{i=1}^{n}p^{q_{i}}(1-2p)^{q-q_{i}}\\
&\times\prod_{j=1,j\neq i}^{n}\left(1-p^{q_{j}}(1-2p)^{q-q_{j}} \right)\\
&=n\phi(p,r)(1-\phi(p,r))^{n-1}.
\end{aligned}
\end{equation}
Now if the configuration $(n,q)$ implies that $\text{Pr}(X^{+}<2)<\frac{1}{2}$ then we can symmetrically assume that $\text{Pr}(X^{-}<2)<\frac{1}{2}$ as well, hence:
$$\text{Pr}(X^{+}\geq 2 \text{ and }X^{-}\geq 2)>0,$$
then the balance has a must-win strategy. 

Again adopting a randomnized strategy to simplify the discussion so:
$$
\text{Pr}(X^{+}<2)=(1-\phi(p,r))^{n-1}\left[1+(n-1)\phi(p,r) \right].
$$

For instance, let use have $r=p=\frac{1}{2}$, then:
$$\phi=2^{-\frac{q}{2}}.$$
When $q$ and $n$ is large, we have:
$$\text{Pr}(X^{+}<2)\approx 1-(n-1)^{2}\phi^{2}=1-(n-1)^{2}2^{-q}.$$
Thus $\text{Pr}(X^{+}<2)<\frac{1}{2}$ is tantamount to:
$$(n-1)^{2}>2^{q-1},$$
a bound of the similar form of that yield by the entropy $n(n-1)>3^{q}$. 

One should note that this bound is significantly lower than that of the entropy-based bound since the entropy rate goes to 2 from 3. A result from $p=\frac{1}{2},r=1$. However, optimize $\text{Pr}(X^{+}<2)$ and $\phi$ w.r.t. $p$ and $r$ is more complex and an analytic solution is hard to find. Generally, the balance tries to minimize $\text{Pr}(X^{+}<2)$ w.r.t. $p$ while the human player tries to maximize $\text{Pr}(X^{+}<2)$ w.r.t. $r$. During which process the approximations made in \eqref{equation:5} and \eqref{equation:6} might fail and a concrete solution is not available. In more general cases where the transcription table is different, the same difficulty appears. Consider two counterfeit coins with $+Q$. One can yield that if $p=\frac{1}{3}$ and $r=\frac{2}{3}$ then if
$$n>2\left(\frac{3}{2} \right)^{\frac{q}{3}}$$ 
then the balance has a must-win strategy.

\subsection{Generalization case: multi-armed electromagnetic balance}
Figure. \ref{figure:0} gives an example of a two-armed electromagnetic balance. What about multi-armed electromagnetic balance? In traditional cases, it is always assumed that a multi-armed balance is capable of addressing one direction to be heavier/lighter out of all its arms. Does such a device exist in reality? 

If the number of counterfeit charged coin is one then it is straightforward to design a $t$-armed balance. Let $t'$ be the smallest odd number that is no smaller than $t$, divide a circle into $t'$ equally curves, putting $t$ boxes onto the mid-points of $t$ different curves will meet the need. The trick here is to make sure that the test charge does not fall onto the connection line between two boxes. 

What if there are two counterfeit charged coin with $+Q$ and $-Q$? One good suggestion for you is to independently and uniformly sample $t$ positions for boxes along the circle centered at the test charge. 

After all, any seemingly \emph{insignificant} assumption or generalization might potentially result in a sharp difference in methodology and reasoning. This line of applying probabilistic method to the balance game is a mere evidence or corollary of this idea.
\bibliographystyle{ieeetr}
\bibliography{balance.bib}

\end{spacing}
\end{document}